# Analysis and Comparison of the LDPC and Reed Solomon Encodings in Mitigating the Imapct of Clipping Noise in OFDM-based VLC

Nima Taherkhani
The University of Texas at Dallas
Richardson, TX 75093
nima.taherkhani@utdallas.edu

Merve Apalak
The University of Texas at Dallas
Richardson, TX 75093
mxa167630@utdallas.edu

Kamran Kiasaleh
The University of Texas at Dallas
Richardson, TX 75093
kamran@utdallas.edu

The linear error-correcting codes are known to be well suited for battling and correcting the burst errors caused by noise in the wireless data transmission system. However, different types of codes offer different decoding and burst-error-correcting capabilities. This paper compares the Low-Density-Parity Check (LDPC) and Reed Solomon (RS) encoding schemes in battling and correcting the burst error caused by the clipping distortion occurred due to the dynamic range constraints in an Orthogonal Frequency Division Multiplexing (OFDM) based Visible Light Communication (VLC). The unipolar conversion applied to the output of the multiplexer in this system results in a clipping noise which distorts the data symbols on the subcarriers in OFDM block. Considering that such distortion impacts the data symbol on each subcarrier differently, RS and LDP are used to encode the data block before being modulated for mapping the OFDM block. In order to control the extreme value of the output of the multiplexer, the transmitter applies puncturing to the generated codeword before mapping OFDM subcarriers, leaving the corresponding subcarriers of the punctured symbols empty. This will lead to the reduction of clipping events in the optical front-end and will mitigate the impact of nonlinear distortion on the modulated symbols for the occupied subcarriers. The redundancy in the codeword generated by the encoder is used not only to control the clipping probability by shortening the number of active subcarriers but also for the reconstruction of the original codeword and correction of the errors caused by channel noise. This work investigates the ability of LDCP and RS encoders in battling the effects of clipping noise in the frequency domain and compares their performances in improving the bit error ratio (BER) performance of an OFDM-based VLC.

## I. INTRODUCTION

Radio Frequency (RF) networks take an important place in our daily lives. According to the mobile data traffic analysis, as the demand for data usage increases, the spectrum used by RF communication will become heavily congested. Therefore, a complementary technology to RF communication will be required. Visible light communication is introduced as a developing technology which has many advantages over RF communication. The development of high-speed LEDs has allowed the realization of VLC systems that can be competitive to RF wireless in short range communications, such as indoor venues, which are primarily connected via Wi-Fi technology at the present time. The first fundamental study of VLC was accomplished by S. Haruyama and M. Nakagawa. They put forward a concurrent illumination and communication system using white LEDs [3]. Komine et al. [4] discussed narrowband OFDM implementation though white LEDs which creates a path for the recent VLC research. In general, VLC systems use LED intensity to carry the data information wirelessly, where the transmitting signal used for intensity modulation has to remain real-valued and positive. In addition, OFDM signals suffer from large peak to average power ratio. it is expected that the transmitting LED will clip the OFDM signal. This fact combined with the dynamic range constraints of LEDs results in double-sided clipping of the OFDM signals. Clipped OFDM signals show optical power efficiency rather than conventional OFDM signals as the range of the signal is controlled by LED. However, the clipping process distorts the transmitting signal and results in performance degradation. Hence, mitigating the impact of nonlinear distortion has become the focus of some studies, for example see [1].

OFDM-based techniques proposed in the literature for generating a real, non-negative VLC signal are limited to two classes; (1) a DC clipped optical-OFDM (DCO-OFDM) [2], and (2) asymmetrical clipped optical-OFDM (ACO-OFDM) [3]. In both techniques, Hermitian symmetry is applied to the symbols of active subcarriers. In ACO-OFDM, data are only mapped on the odd subcarriers and the even subcarriers are set to zero, while in DCO-OFDM, both odd and even subcarriers are employed to carry data. The resulting bipolar signal is added by a DC bias to generate a non-negative signal. For a large constellation, a high-value bias signal might be required to make signal positive. However, to avoid a high peak to average ration (PAPR), a moderate DC bias is added first and then the negative part is clipped at zero. However, due to the limitation of LED's operating range, the time domain signal at the output of the multiplexer is likely to be clipped at both upper and lower bounds of the clipping range. Recently, an alternative approach based on employing Reed Solomon (RS) coding for OFDM-based VLC is proposed to mitigate the impact of clipping noise [4]. It is shown that the reduction of the clipping events at the optical front-end of the transmitter by applying puncturing on the OFDM subcarriers enhances the system bit error rate (BER) performance.

This technique makes use of the block code puncturing feature and erasure decoding capability to control the clipping probability by reducing the magnitude of the time domain signal generated at the output of the IFFT block. Furthermore, the clipping of the transmitting signal at the transmitter may result in burst errors at the output of the de-multiplexer at the receiver side. In systems where error control coding is utilized, minimizing the SNR required to achieve a specific error rate is the ultimate goal. Randomly punctured LDPC codes explained in [5] provides the system with high code rate while transmitting fewer code symbols. It is also proved that randomly punctured LDPC codes have asymptotically good performance when minimum distance analysis is conducted. This work concentrates on analysis and comparison of the LDPC and RS encoding schemes for DCO-OFDM VLC. We compare the capability of two linear block codes in improving the bit error ratio for the AWGN channel under minimum and maximum optical front-end constraints. The effect of codeword puncturing in coded DCO-OFDM VLC is investigated for both single-sided and double-sided clipping scenarios, and the results are given for punctured and regular LDPC coded DCO-OFDM are compared to those of punctured and regular RS coded DCO-OFDM.

The organization of this paper is as follow. In section II, the system model of DCO-OFDM and its formulation are given. In section III, the encoding scheme featured by puncturing process for the DCO-OFDM system is presented. The simulation results are provided in section IV. The concluding remarks are given in section V

## II. System Model of DC Biased Optical OFDM Based VLC

DCO-OFDM is a form of OFDM that modulates the intensity of an LED. Hence, the time domain transmitting signal must be real and non-negative. The block diagram of a DCO-OFDM is presented in Fig.1. The data stream is first mapped onto an M-QAM modulator where the data symbol $X = [X_1, X_2, ..., X_{\frac{N}{2}-1}]$ is generated.

To ensure a real output signal, the Hermitian symmetry is applied to form the complex symbol vector $S$ given by:

$$S = [0, X_1, X_2, ..., X_{\frac{N}{2}-1}, 0, X^*_{\frac{N}{2}-1}, X^*_{\frac{N}{2}-1}, ..., X^*_1]$$

Then the inverse Fourier transform is applied to $I$ to generate a time domain signal. A cyclic prefix is also added in order to turn the linear convolution with channel impulse response into a circular one to eliminate the channel dispersion effect.

For OFDM block with a large number of subcarriers, i.e., $N > 64$ the output of the multiplexer is known to be Gaussian distributed with zero mean and standard deviation $\sigma_i$, i.e., $x_n \sim N(0, \sigma_i^2)$.

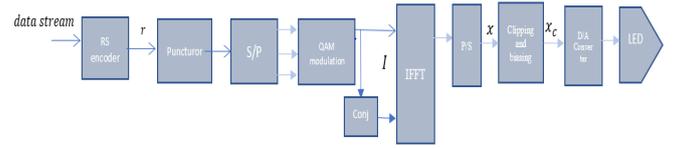

Fig.1 DCO-OFDM transmitter

Since $I_n$ is to be transmitter trough a LED, the unipolar conversion process is applied to transform $I_n$ to a unipolar signal. The unipolar conversion process for DCO-OFDM system entails biasing and clipping of the IFFT signal according to the dynamic operating range of LED, $[I_L, I_H]$:

$$I_{DCO,n} = I_{c,n} + I_{bias} \qquad (1)$$

Where

$$I_{c,n} = C(I_n) \qquad (2)$$

$$= \begin{cases} -\beta_{lower} & I_n \leq -\beta_{lower} \\ I_n & -\beta_{lower} < I_n < \beta_{upper} \\ \beta_{upper} & I_n \geq \beta_{upper} \end{cases}$$

In the above, $C(.)$ denotes the clipping process, $\beta_{upper}$ and $\beta_{upper}$ are the upper lower bounds of clipping, respectively, and $I_{bias}$ is the biasing level which is determined such that the transmitting single satisfies $-I_L < I_{DCO,n} < I_H$. Furthermore, clipping bounds are usually set relative to the standard deviation of $I_n$ using a proportionally constant $\alpha_1$ and $\alpha_2$ [6]:

$$\beta_{lower} = \alpha_1 \sqrt{E\{I_n^2\}} \qquad (3)$$
$$\beta_{upper} = \alpha_2 \sqrt{E\{I_n^2\}}$$

where $E\{.\}$ denoted the expectation of the enclosed. Generally, in literature the clipping process is modelled using Bussgang theorem [7]:

$$I_{n,c} = \gamma I_n + \varepsilon_n \qquad (4)$$

where $\gamma$ is the attenuation factor and $\varepsilon_n$ is the clipping noise, which is assumed to be uncorrelated with signal $I_n$ [8].

The attenuation factor is also given by $a = \frac{E\{I_{c,n} I_n\}}{E\{I_n\}}$. Considering a baseband linear time-invariant VLC channel, the received signal is given by:

$$y_n = (\gamma I_n + \varepsilon_n + I_{bias}) * h_n + w_n, \qquad (5)$$

where $h_n$ is the channel impulse response, summation of the channel and shot noise is given by $w_n$ which is modelled as a Gaussian noise with zero-mean and variance $\sigma_w^2$, and * denotes the convolution operation. Provided that the DC biasing level contributes only to the $0^{th}$ subcarrier, the signal received at $k$ subcarrier in frequency domain signal is given by:

$$Y_k = \gamma H_k S_k + H_k F_k + W_k, \quad 1 \leq k \leq N/2 \quad (6)$$

where $H_k$ and $F_k$ denotes the channel frequency component for the $k^{th}$ subcarrier and the $k^{th}$ term of the Fourier transform of the clipping noise, respectively. Different works have studied the impact of $\gamma$ and $D_k$ on the performance of the DCO-OFDM which employ linear equalizers for the detection and estimation of the transmitted signal using the linear model given by (5). However, this approach can lead to a sub-optimal solution for those circumstances where linear modelling does not apply to the clipping distortion as demonstrated in [4]. In the following section, two linear block codes, *i.e.*, Reed Solomon and LDPC codes, are studied side by side for the performance enhancement of DCO-OFDM VLC communications in the presence of non-negligible clipping noise for different clipping circumstances. The motivation behind this approach stems from the fact that, increasing the power of transmitting signal in a DCO-OFDM system which utilizes an LED with a short and bounded transmittable range not only does not improve the throughput but also leads to an even larger clipping noise, which can degrade its BER performance. Using linear block codes for encoding the data stream of DCO-OFDM and employing the puncturing feature can help reduce the burst errors caused by clipping and also will aid in correction and reconstruction of the transmitted data using the redundancy in the encoded message.

## III. SYSTEM MODEL OF CODED DCO-OFDM

The method used in this work for mitigating the clipping noise incorporates data block expansion and puncturing using linear block code, such as LDPC and Reed Solomon codes. RS codes are maximum distance separable (MDS), which means they achieve the maximum possible distance, $D_{min}$, for a forward error correction (FEC) coding scheme. LDPC, as an alternative to RS coding, offers a low decoding complexity and parallel implementation. The structure of the transmitter for a coded DCO-OFDM is given in Fig.1. The incoming bits are either encoded by an LDPC encoder or converted to decimal and fed to the RS encoder to generate the codeword as the input for the M-QAM modulator. The codeword $c$ at the output of the encoder is subjected to a different number of puncture. The choice of the number $v_c$ of the punctured elements in the codeword depends on the tradeoff between bit error rate (BER) reduction and spectral efficiency. Punctured data vector is then applied to M-QAM and then QAM symbols will perform the OFDM frame mapping. The subcarriers corresponding to the punctured elements in the codeword are set to zero in order to reduce the clipping probability by mapping zero symbols on subcarriers corresponding to the punctured elements. For LDPC code, a greedy upper triangulation-based encoder can be used which is demonstrated as the efficient and manageable encoding algorithm for LDPC code [32 of theist]. According to [32], instead of finding a generator matrix, H can be directly used to transform it into the desired upper triangular form depicted in Figure 3.3 of [32]. The idea under the proposed encoding procedure is to keep the parity check matrix as sparse as possible by row and column permutations. Encoding can be separated into two main steps which are preprocessing and encoding steps. According to [32], instead of finding generator matrix $G$, the parity check matrix $H$ can be directly used to transform it into the desired upper triangular form. The idea under the proposed encoding procedure is to keep the parity check matrix as sparse as possible by row and column permutations [32]. At the receiver side, the M-QAM demodulator generates the Log-Likelihood-Ratio (LLR) value for each demodulated bit which is then used by the decoder to perform soft decoding. The Sum-Product Algorithm (SPA) is employed by the decoder to calculate the maximum a posteriori (MAP) estimations of input bits. Considering that the decoder has no knowledge of the punctured symbols in the original codeword, it assigns zero LLR value for the indices corresponding to the punctured symbols. Here, the number of punctured $v_s$ plays an important role in controlling the variation of time domain signal generated by the multiplexer. If only a small number of symbols are punctured, then most of the redundancy is retained in the codeword, so they can be used for error correction at the receiver. However, in that case, a larger number of OFDM subcarrier will be active which will generate a time domain transmitting signal with a larger range in value. This can make it more likely for the signal to get clipped during the unipolar conversion process. Having a larger number of symbols getting punctured, the probability that the time domain signal exceeds the clipping range becomes smaller. However, in such a case, the error correcting capability of the codeword is compromisedIn RS encoding scheme, an RS $(N', K')$ encoder is to be employed for encoding of the incoming bits. Number of correctable errors of each codeword in this scheme with $v_s$ punctured symbols is given by $t_m = \left\lfloor \frac{N'-K'-v_s}{2} \right\rfloor$. At the receiver side, an eraser decoder is employed to reconstruct and correct the original transmitted codeword, provided that the receiver has been informed about the indices of the punctured symbols in the codeword in advance. Here again, $v_s$ plays an important role in establishing a trade-off between error correction capabilities of the decoder and mitigating the clipping noise impact by controlling the clipping events.

## IV. SIMULATION RESULTS

The potential of LDPC and Reed-Solomon encodings in mitigating the detrimental impact of clipping noise in DCO-OFDM systems is assessed using computer simulation. Reed-Solomon and LDPC coding-based schemes are incorporated with OFDM for message block expansion and puncturing in order to compensate non-linear distortion caused by the double-sided clipping in the DCO-OFDM system. Both coding schemes allow the system to reduce the magnitude of the transmitted signal and to regulate clipping process at the transmitter by the use of the puncturing approach. LDPC, which is well-known for its error correction potential, is compared to Reed Solomon coding, which is known as a distinctive scheme in reaching the maximum possible distance. The performance of the LDPC and Reed-Solomon encoded DCO-OFDM systems in the presence of the double-sided clipping employment are assessed in terms

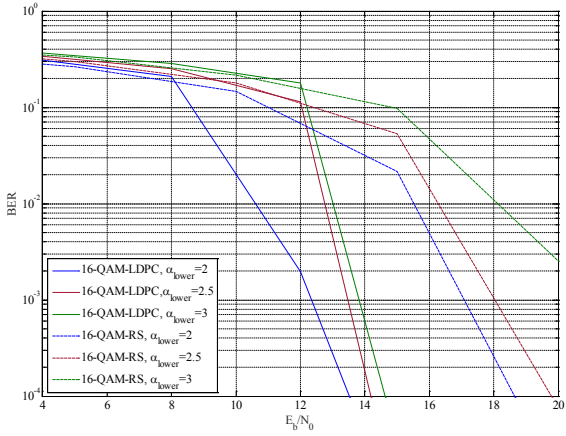

Fig. 2: Bit Error Ratio performance of LDPC encoding DCO-OFDM vs RS encoding DCO-OFDM in AWGN channel assuming 16-QAM employment for different clipping ranges.

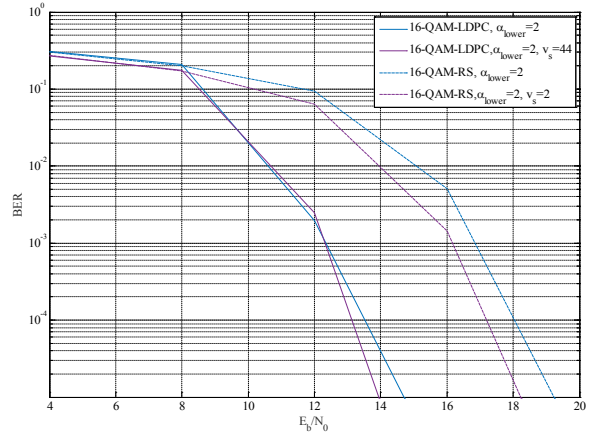

Fig. 3: Bit Error Ratio performance of LDPC encoding DCO-OFDM vs RS encoding DCO-OFDM with 16-QAM modulator in AWGN channel when punctured codewords are used for OFDM block mapping for $\alpha_{lower} = 2$

of BER which is defined as a function of the electrical energy to noise power ratio per useful bit $\frac{E_b}{N_0}$. We also assume an LED with a dynamic range of $[0, 5\sigma_I]$, which yields the clipping and biasing parameters as: $\beta_{lower} = \alpha_1\sigma_I$, $\beta_{upper} = (5 - \alpha_2)\sigma_I$, and $I_{bias} = \alpha_1\sigma_I$. At the receiver side, we assume a photodiode with unit optical power. Considering that the received signal is first de-biased at receiver, the power of useful signal is normalized using the attenuation factor given by : $a_{bias} = \frac{\sigma_I^2}{\sigma_I^2 + I_{bias}^2}$. We consider a system with $N$=512 subcarriers, where 16-QAM modulation is used to generate symbols for OFDM block mapping. Considering a DCO-OFDM system, $N_f = \frac{512}{2} - 1 = 255$ subcarriers are available for data transmission. For LDPC encoding scheme, a half rate encoder with the input data vector of length 510 is employed to generate a codeword of the length 1020. The codeword is then forwarded to the 16-QAM modulator for OFDM symbol generation. A DCO-OFDM system with the same number of subcarriers is also considered for Reed-Solomon encoding scheme, where an RS (15,8) encoder is employed in order to generate a similar data rate for fair comparison. For this scheme, again a binary data vector of length 1020 is used to generate the decimal input vector.

The RS (15, 8) encoder generates 17 codewords with 15 symbols each, which are concatenated for generating a symbol vector of length 255 as the input for the 16-QAM modulator. For both schemes, after applying the Hermitian symmetry and IFFT operation, the output of the multiplexer is clipped according to different clipping bounds in order to explore the optimum lower bound for each scheme.

Fig. 3 shows the performance of LDPC and Reed-Solomon encoding schemes used in DCO-OFDM system where the transmitting signal is subject to the different clipping scenario. As can be seen, for both schemes, the optimum clipping range in the underlying system yields at $\beta_{lower} = 3\sigma_I$. In order to investigate the contribution of the puncturing approach in mitigating the detrimental impact of the clipping distortion, we also compare the BER performance of two encoding schemes when punctured codewords are used for data transmission. In this comparison, the output of each encoder is exposed to different numbers of puncturing which results in a shorter codeword for both schemes. After modulating the codeword symbols, the OFDM subcarriers that correspond to the punctured symbols in the codeword are left empty and inactive. For LDPC encoded message, we puncture $v_s = 44$ symbols in the codeword. For the RS encoder, $v_s = 2$ symbols out of 15 symbols are punctured in each codeword. We also consider the clipping parameters to be $\alpha_{lower} = 2\sigma_I$, $\alpha_{upper} = 3\sigma_I$, which were shown to be the optimum clipping parameters in Fig. 2. Fig. 3 depicts the performance of two schemes when punctured and unpunctured codewords are used for data transmission. As the figure shows, LDPC scheme presents 0.4 dB gain at the target BER of $10^{-4}$, while the improvement for Reed-Solomon scheme at the same target BER is 0.75 dB.

It should be noted that the numbers of puncturing in two encoding scheme are adjusted in a way that a similar date rate is achieved for both schemes. We also compare the performance of two schemes when a higher constellation order, *i.e.*, 64-QAM, is employed. We consider again as system with $N$=512 subcarriers, where 6 subcarriers are used for header and prefix insertion, whereby $N_f = \frac{506}{2} - 1$ of them are to be used by DCO-OFDM for data transmission. For LDPC encoding scheme, again an encoder with half rate is employed, which receives a data vector of length 756 to generate the codeword of the length 1530. The codeword is then applied to a binary to decimal converter whose output is then used to generate the symbols for mapping to the 252 OFDM subcarriers. For Reed-Solomon encoding scheme, an RS (63, 31) is employed which generates and concatenates four consecutive codewords with the size of 63 symbols by using the incoming bits. After mapping the OFDM block and applying the Hermitian symmetry, IFFT operation is conducted to generate the real-valued time domain signal. The bipolar signal is converted to its unipolar version using the parameters specified in this section. Fig. 4 shows the performance of two scheme for different clipping scenarios. As can be seen, the optimum lower clipping bound for both schemes is attained at $\beta_{lower}=2.5\sigma_I$. In another comparison (see Fig. 5), the puncturing approach is taken again to examine the enhancement it can provide in the DCO-OFDM system using two mentioned encoding schemes when 64-QAM modulation is employed. We set the unipolar conversion parameters as $I_{bias}=\beta_{lower}=2.5\sigma_I$, $\beta_{upper}=2.5\sigma_I$ which were attained as the optimum values for the system of interest. For the sake of fair comparison, we adjust the number of punctured symbols in LDPC and RS encoded messages such that a similar data rate is achieved for both schemes. For RS scheme, 5 and 20 symbols are punctured in each codeword in separate scenarios. These puncturing processes will result in 20 and 80 subcarriers to be left empty and inactive in the OFDM block, respectively. For LDPC encoding scheme, the last 20 and 80 symbols of the codewords are punctured in order to achieve the same data rate in each corresponding scenario. Fig 5 shows the performance of two schemes when the puncturing process is performed. It can be seen that RS coded DCO-OFDM is better than its unpunctured counterpart at the target BER of $10^{-4}$ by almost $3.7\ dB$ when $v_s=20$. LPDC encoding scheme also offers several dB of performance enhancements when $v_s=80$ as compared to the 20 punctured symbols or no puncturing scenarios. It can also be seen that for higher modulation order, RS encoding scheme offers a superior performance as compared to the LDPC encoding scheme used in DCO-OFDM system.

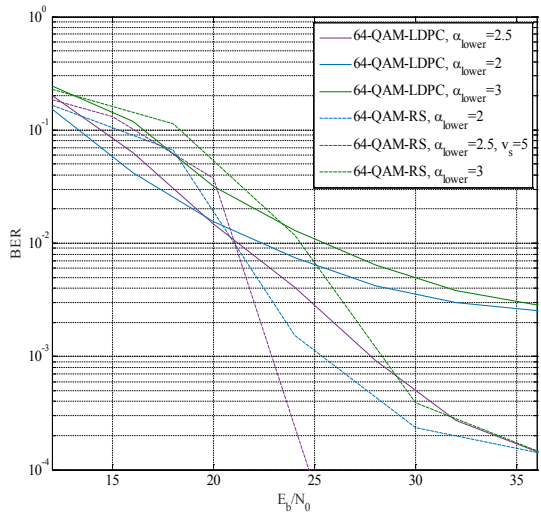

Fig. 4: Bit Error Ratio performance of LDPC encoding DCO-OFDM vs RS encoding DCO-OFDM in AWGN channel assuming 64-QAM employment for different clipping ranges

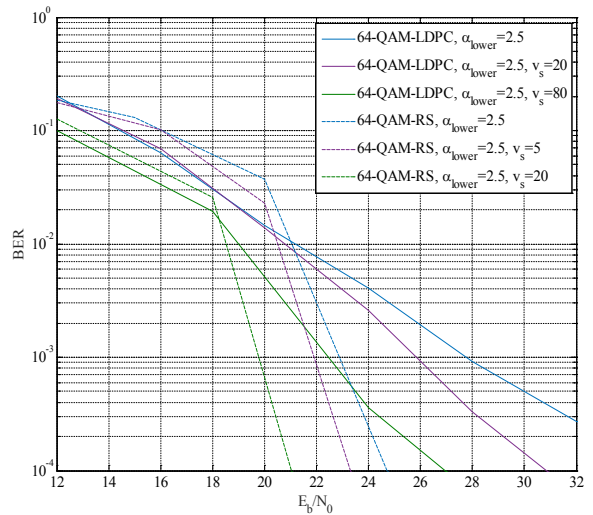

Fig. 5: Bit Error Ratio performance of LDPC encoding DCO-OFDM vs RS encoding DCO-OFDM with 64-QAM modulator in AWGN channel for different number of puncture in the encoded messages for $\alpha_{lower}=2.5$

## V. CONCLUSION

In this work, we compared the performances of two linear block codes in battling the impact of the clipping noise in OFDM-based VLC. We showed that the impact of nonlinear distortion caused by the dynamic range constrains of the optical front-end can be mitigated with the aid of burst error tolerance and puncturing feature of LDPC and Reed-Solomon coding. Although the integration of block coding schemes with OFDM-based VLC results in the overall data rate reduction, the puncturing feature of linear codes can be employed to reduce the clipping probability in VLC by controlling the extreme values of the OFDM signal. Furthermore, the advantage of integrating encoding schemes with DCO-OFDM enables the correction of the burst error caused by the clipping noise of the OFDM symbols. Results presented in this work show that the application of punctures to the encoded message leads to a better BER performance for both encoding schemes compared to the no puncturing scenario. Whilst LDPC encoding scheme integrated with DCO-OFDM is more robust against nonlinear distortion when lower level modulation constellation, *i.e.*, 16-QAM, is considered, the RS encoding scheme offers a superior BER performance when compared with its LPDC counterpart for higher modulation orders.